\documentclass[]{spie}  

\usepackage{amsmath,amsfonts,amssymb}
\usepackage{graphicx}
\usepackage{float}
\usepackage[colorlinks=true, allcolors=blue]{hyperref}

\title{A Wideband Millimeter-wave Receiver at 210-350 GHz for LMT-FINER}

\author[a]{Haoran Kang}
\author[a]{Takafumi Kojima}
\author[b]{Takeshi Sakai}
\author[c]{Yoichi Tamura}
\author[a,g]{Shun Ishii}
\author[d]{Akio Taniguchi}
\author[c]{Masato Hagimoto}
\author[c]{Masato Kato}
\author[e]{Taku Nakajima}
\author[a]{Tai Oshima}
\author[a]{Sho Masui}
\author[f]{Issei Watanabe}

\affil[a]{National Astronomical Observatory of Japan, 2-21-1 Osawa, Mitaka, Tokyo 181-8588, Japan}
\affil[b]{The University of Electro-Communications, 1-5-1 Chofugaoka, Chofu, Tokyo 182-8585, Japan}
\affil[c]{Nagoya University, Furocho, Chikusa-ku, Nagoya, Aichi 464-8602, Japan}
\affil[d]{Kitami Institute of Technology, 165 Koen-cho, Kitami, Hokkaido 090-8507, Japan}
\affil[e]{Suwa University of Science, 5000-1 Toyohira, Chino, Nagano 391-0292, Japan}
\affil[f]{National Institute of Information and Communications Technology, 4-2-1 Nukui-Kitamachi, Koganei, Tokyo 184-8795, Japan}
\affil[g]{The Graduate University for Advanced Studies (SOKENDAI), 2-21-1 Osawa, Mitaka, Tokyo
181-8588, Japan}

\authorinfo{Further author information: (Send correspondence to H.K)\\H.K: E-mail: haoran.kang@nao.ac.jp\\}

\begin{document} 
\maketitle

\begin{abstract}
The Far-Infrared Nebular Emission Receiver (FINER) project is developing two wideband dual-polarization sideband-separating receivers covering 120–210 GHz and 210–350 GHz to efficiently identify high-redshift galaxy candidates in the early universe. Based on high-critical-current-density superconductor--insulator--superconductor mixer technology originally developed for the ALMA wideband sensitivity upgrade, the FINER receivers are designed to provide an intermediate-frequency bandwidth of 3–21 GHz per sideband and per polarization, approximately five times wider than the current ALMA specifications. After installation on the Large Millimeter Telescope, these receivers are expected to offer highly efficient spectral-scanning capability among (sub)millimeter-wave facilities in the northern-hemisphere.

This paper reports the initial laboratory characterization of the 210–350 GHz receiver. The measured single-sideband receiver noise temperature is approximately 100 K over most of the radio-frequency band. Digital sideband separation was also demonstrated using a wideband spectrometer array (DRS4), achieving an image rejection ratio of around 20 dB in the initial tests. These results represent an important step toward the realization of a wideband spectral-scanning receiver system for FINER.

\end{abstract}


\section{Receiver Noise Temperature Performance}
\label{sec:Trx}  

\begin{figure} [h]
   \begin{center}
   \begin{tabular}{c} 
   \includegraphics[height=9cm]{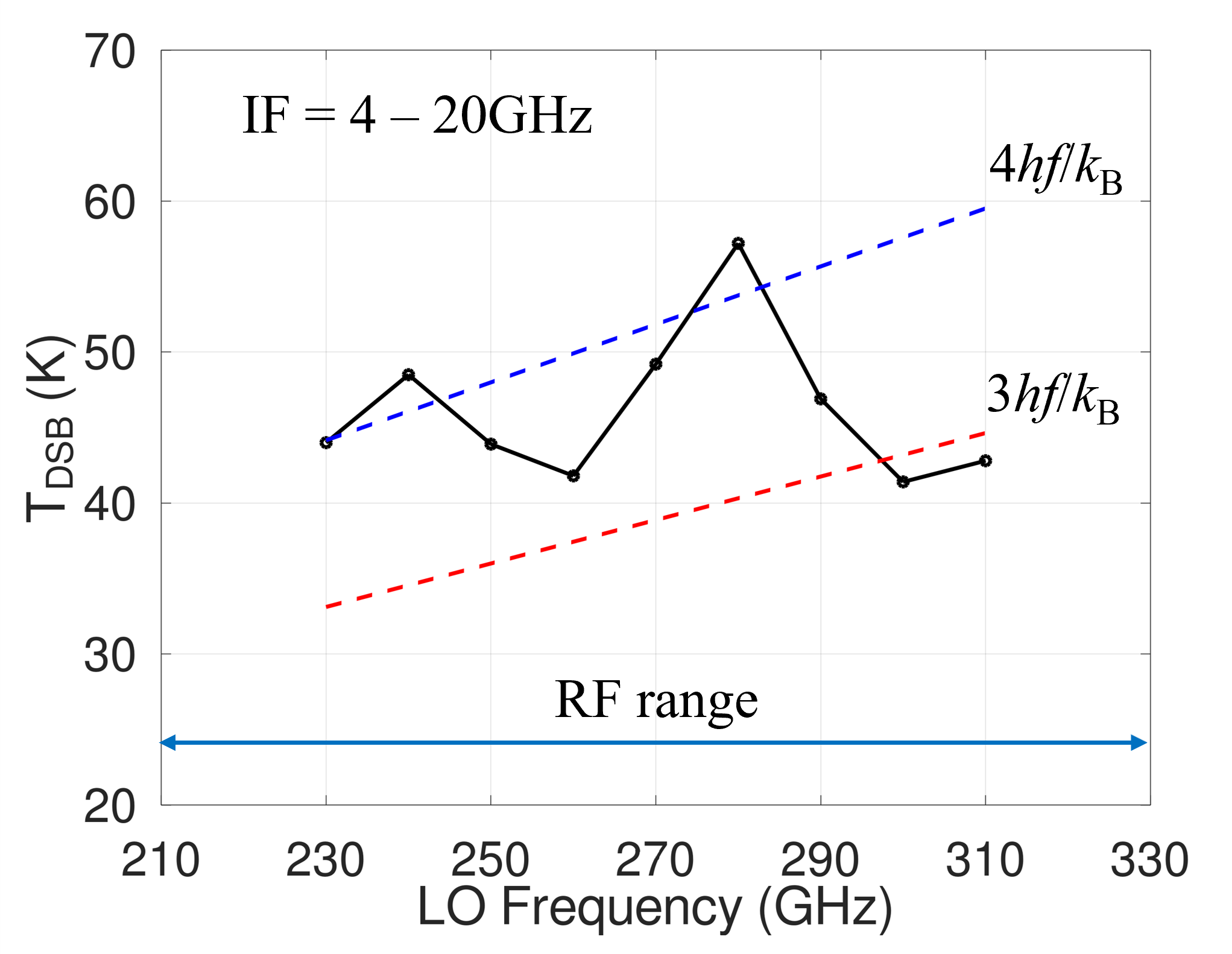}
   \end{tabular}
   \end{center}
   \caption[double-sideband-noise] 
   { \label{fig:T-dsb} 
Measured double-sideband (DSB) receiver noise temperature of the 210–350 GHz mixer for Far-Infrared Nebular Emission Receiver (FINER) as a function of local oscillator (LO) frequency.}
   \end{figure} 

The key wideband waveguide components for the 210--350 GHz receiver were reported at this conference two years ago~\cite{kang2024development}; therefore, their details are not repeated here. The present proceeding focuses on the initial laboratory characterization of this receiver, including the receiver noise temperature and image rejection ratio (IRR). 

For the wideband superconductor--insulator--superconductor (SIS) mixer, we employ high-critical-current-density (high-$J_{\mathrm{c}}$) SIS junction technology. The low resistance of high-$J_{\mathrm{c}}$ SIS junctions enables the radio-frequency (RF) and intermediate-frequency (IF) bandwidths to be extended toward wider frequency coverage~\cite{kerr1995some, kojima2017performance, kojima2018275, kojima2020wideband}. We then measured the double-sideband (DSB) receiver noise temperature of the 210--350 GHz mixer. A typical measurement result, averaged over an IF bandwidth of 4--20 GHz, is shown in Fig.~\ref{fig:T-dsb}. The SIS mixer showed an average DSB noise temperature of approximately 40--55 K over an local oscillator (LO) frequency range of 230--310 GHz, corresponding to about three to four times the quantum limit over an RF range of 210--330 GHz. Measurements above an RF frequency of 330 GHz were not performed mainly because of the frequency limitation of the tunable filter in the current LO chain, and it will be updated. 

\begin{figure} [htbp]
   \begin{center}
   \begin{tabular}{c} 
   \includegraphics[height=9cm]{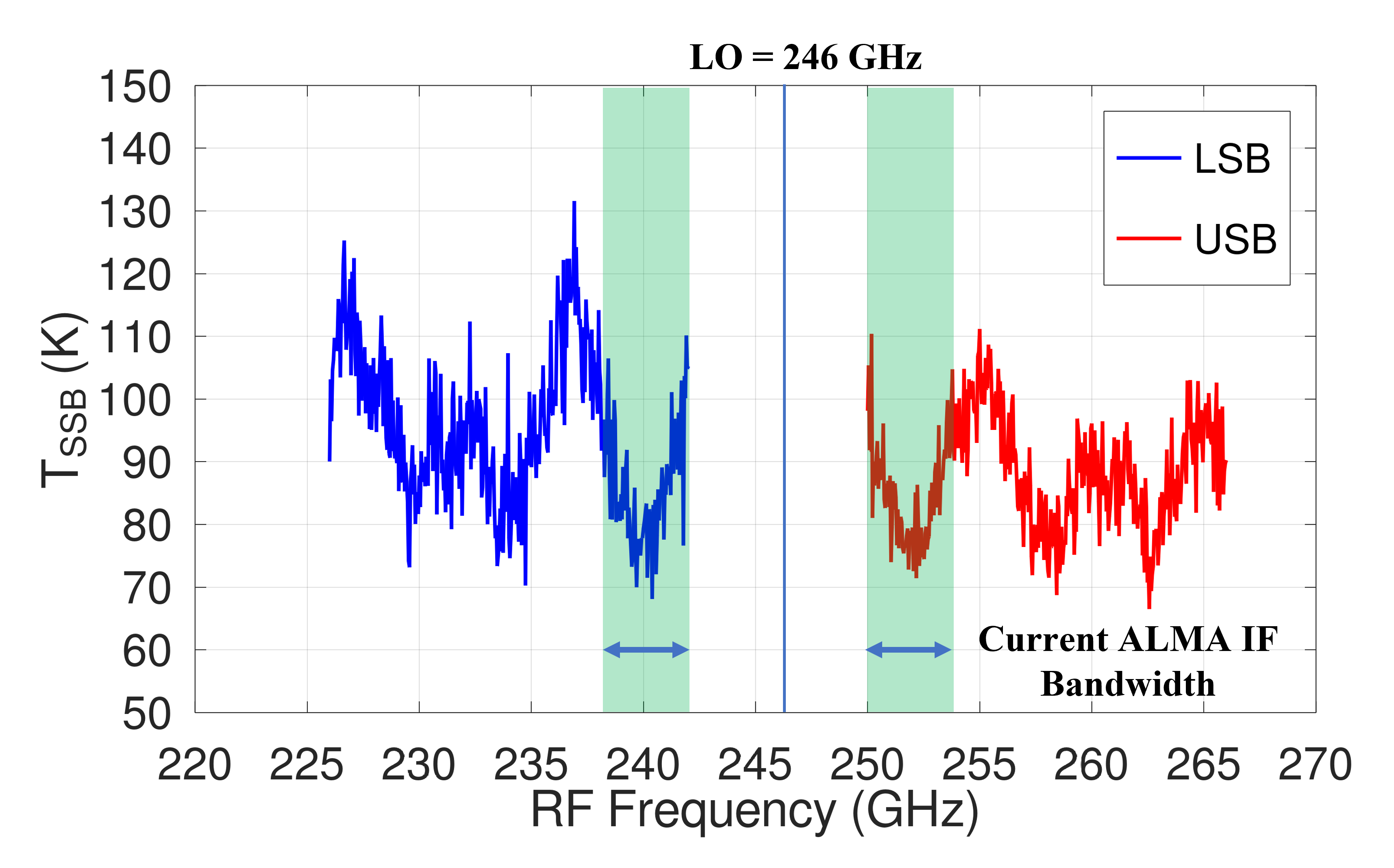}
   \end{tabular}
   \end{center}
   \caption[single-sideband-noise] 
   { \label{fig:T-ssb} 
Representative single-sideband (SSB) receiver noise temperature of the FINER 210--350 GHz receiver system measured at an LO frequency of 246 GHz. The shaded green regions indicate the current ALMA IF bandwidth of 4--8 GHz.}
   \end{figure}

We further evaluated the receiver noise temperature in sideband-separating (2SB) configuration over the same LO frequency range as the DSB measurement, from 230 to 310 GHz. Overall, the measured single-sideband (SSB) receiver noise temperature was approximately 100--120 K, which is consistent with the DSB receiver noise temperature results. Fig.\ref{fig:T-ssb} shows a representative SSB noise temperature result measured at an LO frequency of 246 GHz for single polarization. The blue and red curves correspond to the lower sideband (LSB) and upper sideband (USB) respectively.

The shaded regions in Fig.\ref{fig:T-ssb} indicate the current ALMA IF bandwidth of 4--8 GHz. Within this IF range, the FINER receiver shows an comparable noise temperature of current ALMA receivers, but more importantly, the receiver shows a excellent noise temperature over a much wider IF bandwidth of 4--20 GHz, which is four times wider than the current ALMA IF coverage \cite{kerr2014development}. 
The noise temperature below IF frequencies of 4 GHz is degraded, mainly due to the noise contribution from the current LO chain. The receiver noise temperature above IF frequencies of 20 GHz was not evaluated in this noise temperature measurement, because the spectrum analyzer used for the this measurement was limited to 20 GHz.

Although this receiver is designed for dual-polarization operation, only the results for a single polarization are presented in this proceeding. This is because the present experiment was performed using one wideband spectrometer array (DRS4), which can process the DC--10.24 GHz and 10.24--20.48 GHz IF signals for the two sidebands of a single polarization~\cite{hagimoto202410}. Characterization of the second polarization is currently in progress, and the results will be reported in a future publication.

\begin{figure} [htbp]
   \begin{center}
   \begin{tabular}{c} 
   \includegraphics[height=9cm]{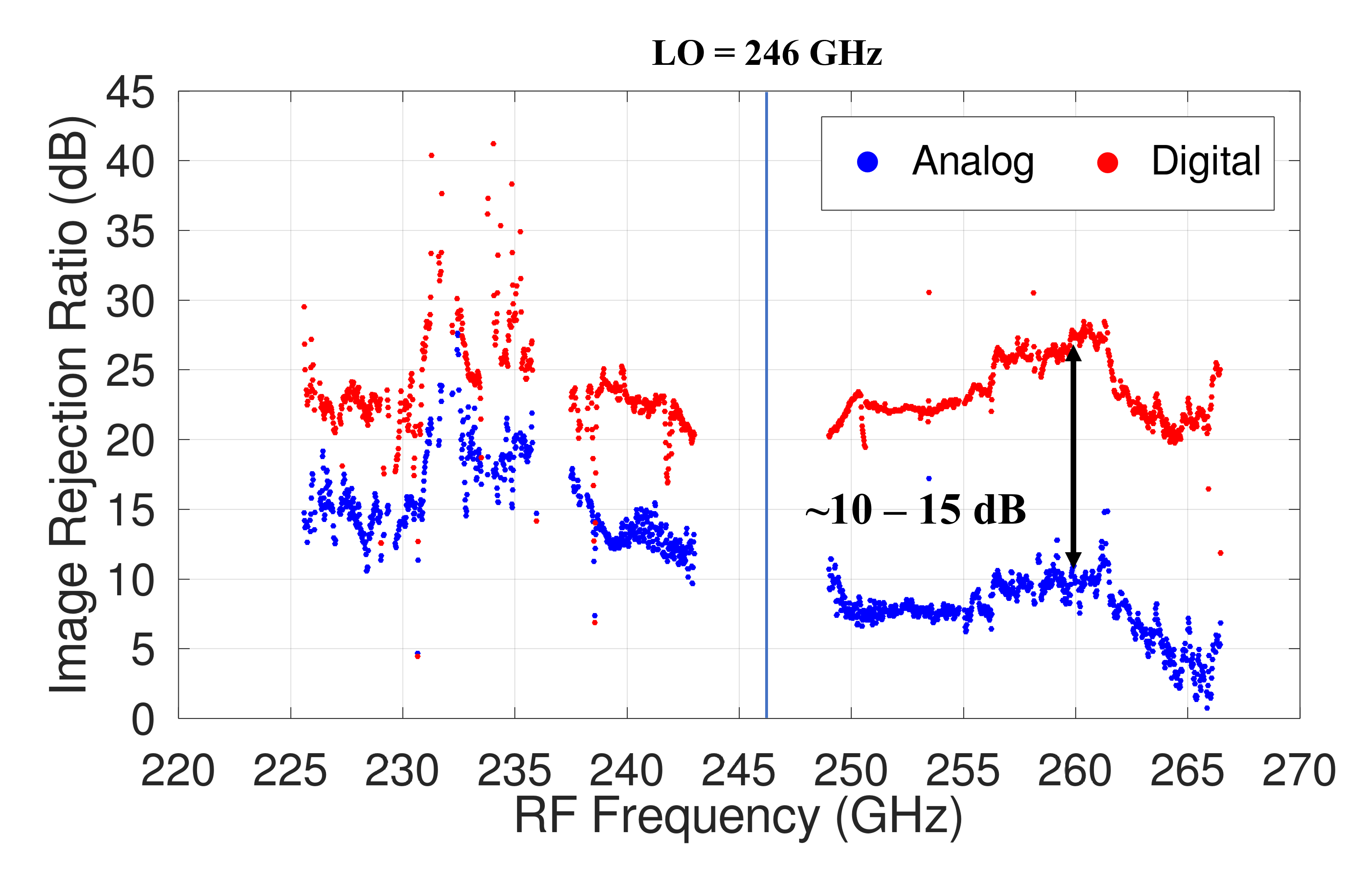}
   \end{tabular}
   \end{center}
   \caption[Image-rejection-ratio] 
   {\label{fig:IRR} 
Image rejection ratio (IRR) of the FINER 210--350 GHz receiver system measured at an LO frequency of 246 GHz. The blue and red points show the IRR before and after applying digital sideband separation, respectively. Channels with a signal-to-noise ratio lower than five in the signal sideband and/or lower than three in the image sideband are excluded from this figure.}
   \end{figure} 

\section{Sideband Separation Performance}

In a 2SB receiver, imperfections in the amplitude and phase balance of the analog RF and IF components cause leakage between the USB and LSB signals, which limits the IRR. Conventional 2SB receivers at this frequency range typically achieve an IRR of approximately 10--15 dB~\cite{kerr2014development}. However, achieving a high IRR becomes more challenging for receivers with wider IF bandwidths, because the amplitude and phase balance must be maintained over a wider frequency range \cite{garrett2023345}.
For the FINER project, both the RF and IF bandwidths are significantly wider than those of existing state-of-the-art receivers. Therefore, achieving the target IRR specification of 25 dB is particularly challenging \cite{tamura2024finer}. A high IRR is important for reducing the atmospheric-noise contribution from the image sideband and for achieving the sensitivity required for the scientific goals of FINER project(details to be presented by Taniguchi et al., this conference, 14156-31).

To improve the sideband separation performance, we use digital sideband separation (DSBS) algorithm with the wideband  spectrometer array (DRS4) developed for FINER project~\cite{hagimoto202410}. In this algorithm, an artificial continuous wave reference signal is injected into the receiver, and the auto-correlation and cross-correlation spectra of the USB and LSB output ports are measured. The complex gains required to cancel the image sideband leakage is then derived for each spectrometer channel and applied in the digital backend \cite{murk2009high, finger2013calibrated, rodriguez2018digital}. Fig.\ref{fig:IRR} shows the measured IRR before and after applying DSBS at an LO frequency of 246 GHz. Before digital correction, the IRR obtained by analog sideband separation was typically around 10 dB and degraded to below 10 dB at some frequencies. After applying DSBS, the IRR was improved to approximately 20 dB or higher over most of the measured frequencies. The IRR increased by approximately 10--15 dB compared with the analog result in general. These results demonstrate that the FINER receiver, in combination with the DSBS, can significantly improve the sideband separation performance required for wideband astronomical observations.

\section{CONCLUSION}

We presented the initial laboratory characterization of the FINER 210--350 GHz wideband receiver. The measured DSB receiver noise temperature was approximately 40--55 K over an LO frequency range of 230--310 GHz, and the SSB receiver noise temperature was approximately 100--120 K over most of the measured RF band. These results demonstrate low-noise receiver performance over a wide IF bandwidth of 4--20 GHz.

Digital sideband separation was also demonstrated using the wideband spectrometer array. The IRR was improved to approximately 20 dB or higher over most of the measured frequency range. Further characterization of the second polarization is currently in progress.

\acknowledgments
This work was financially supported by JSPS KAKENHI (Nos. 22H04939, 22J21948, 22KJ1598, 23K20035), NAOJ Joint Development Research Programs (Nos. 1901-0101, 2001-0104).
We acknowledge the use of the facilities at the Advanced Technology Center (ATC) of the National Astronomical Observatory of Japan (NAOJ).
The LMT project is a joint effort between the Instituto Nacional de Astr\'{o}fisica, \'{O}ptica, y Electr\'{o}nica (INAOE) and the University of Massachusetts at Amherst (UMASS).

\bibliography{report} 
\bibliographystyle{spiebib} 
\end{document}